\documentclass[dvipdfmx]{pasj01}

\begin{document} 
\Received{2017/12/28}
\Accepted{2018/2/6}

\title{Triggering the formation of the supergiant H\emissiontype{II} region NGC 604 in M33}

\author{Kengo \textsc{Tachihara}\altaffilmark{1}}%
\altaffiltext{1}{Department of Physics, Nagoya University, Furo-cho, Chikusa-ku, Nagoya, 
Aichi 464-8602, Japan}
\email{k.tachihara@a.phys.nagoya-u.ac.jp}

\author{Pierre \textsc{Gratier},\altaffilmark{2}}
\altaffiltext{2}{Laboratoire d'Astrophysique de Bordeaux, Univ. Bordeaux, CNRS, B18N, All{\'e}e Geoffroy Saint-Hilaire, 33615 Pessac, France}

\author{Hidetoshi \textsc{Sano}\altaffilmark{1,3}}
\altaffiltext{3}{Institute for Advanced Research, Nagoya University, Furo-cho, Chikusa-ku, Nagoya 464-8601, Japan}

\author{Kisetsu \textsc{Tsuge}\altaffilmark{1}}

\author{Rie \textsc{E.~Miura},\altaffilmark{4}}
\altaffiltext{4}{Chile Observatory, National Astronomical Observatory of Japan, 
National Institutes of Natural Sciences,2-21-1 Osawa, Mitaka, Tokyo 181-8588, Japan}

\author{Kazuyuki \textsc{Muraoka},\altaffilmark{5}}
\altaffiltext{5}{Department of Physical Science, Graduate School of Science, 
Osaka Prefecture University, 1-1 Gakuen-cho, Naka-ku, Sakai, Osaka 599-8531, Japan}

\author{Yasuo \textsc{Fukui}\altaffilmark{1,3}}

\KeyWords{ISM: clouds --- ISM: individual objects (M33) --- stars: formation --- radio lines: ISM} 

\maketitle

\begin{abstract}
Formation mechanism of a supergiant H\emissiontype{II} region NGC 604 is discussed 
in terms of collision of H\emissiontype{I} clouds in M33. An analysis of the archival 
H\emissiontype{I} data obtained with the Very Large Array (VLA) reveals complex velocity 
distributions around NGC 604. The H\emissiontype{I} clouds are composed of two velocity 
components separated by $\sim 20$ km s$^{-1}$ for an extent of $\sim 700$ pc, beyond 
the size of the the H\emissiontype{II} region. Although the H\emissiontype{I} clouds are 
not easily separated in velocity with some mixed component represented by merged line 
profiles, the atomic gas mass amounts to $6 \times 10^6~M_{\Sol}$  and $9 \times 
10^6~M_{\Sol}$ for each component. 
These characteristics of H\emissiontype{I} gas and the distributions of dense molecular 
gas in the overlapping regions of the two velocity components suggest that the formation 
of giant molecular clouds and the following massive cluster formation have been 
induced by the collision of H\emissiontype{I} clouds with different velocities. 
Referring to the existence of gas bridging feature connecting M33 with M31 reported by 
large-scale H\emissiontype{I} surveys, the disturbed atomic gas possibly represent 
the result of past tidal interaction between the two galaxies, which is analogous to the 
formation of the R136 cluster in the LMC. 

\end{abstract}

\section{Introduction}

Dynamical interactions of galaxies are believed to be a major driver of active star formation. 
Large scale galaxy collision or merger make active starbursts happen, and accelerate the 
galaxy evolution and metal enrichment (e.g., \cite{genzel98}). 
It is widely believed that galaxies evolve as they undergo collisions to merge with other galaxies. 
Physical processes of the massive cluster formation in detail has not been well 
understood yet since these are not spatially resolved due to their large distances.
The Milky Way Galaxy should also have experienced such dynamical events in the past although 
the current star formation activity is not so high as those observed in starburst galaxies. 
The existence of globular clusters in the Galactic halo suggests past starburst activities, and 
hence, investigation of the formation of rich star cluster is a key to understand the evolution 
history of the Galaxy. 

Close encounters of galaxies, on the other hand, as a small-scale interaction between galaxies 
is supposed to perturbs materials in galaxies and enhance the star formation activity 
(e.g., \cite{noguchi86}). The tidal force acting between two galaxies during the encounter strips 
off portion of the interstellar material, and the material may fall back down onto the galaxies 
after the encounter. The Magellanic system is suggested to be one example of such events 
as theoretically proposed by \citet{fujimoto90}, and represented with numerical simulations by 
\citet{bekki07a, bekki07b}. These simulations show that the two Magellanic clouds had 
close encounters in about $2 \times 10^8$ yr ago and streaming H \emissiontype{I} gas 
is falling down onto the disk of the Large Magellanic Cloud (LMC) with a velocity of $\sim 50$ 
km s$^{-1}$. Most recently, \citet{fukui17} suggested that the super star cluster RMC 136 
(hereafter R136) has been formed by triggering of colliding H \emissiontype{I} flow falling 
from the Small Magellanic Cloud (SMC) to the LMC. 

M33 is a member of the local galaxy group associated with M31, and harbors NGC 604 that is 
a super star cluster similar system to R136. 
H \emissiontype{I} high velocity clouds (HVCs) in the proximity of M33 have been discovered, 
and possible origins are discussed \citep{grossi08}. Among them, gas fueling from M31 
seems to be reasonable since the bridge of H \emissiontype{I} gas, called M31-M33 stream, 
is detected by the Westerbork Sythesis Radio Telescope (WSRT) \citep{braun04}, the Arecibo 
radio telescope \citep{putman09}, and Green Bank Telescope (GBT) \citep{lockman12}. 
The M31-M33 stream is successfully reproduced by numerical simulations similar to 
the Magellanic stream \citep{bekki08}. 
These are thus compelling evidence that continuous gas accretion of 0.5 $M_{\Sol}$ yr$^{-1}$ 
provides necessary fueling for sustaining M33's relatively active star formation rate 
\citep{grossi08}. This implies that formation of young super star cluster NGC 604 in M33, the 
largest and brightest H \emissiontype{II} region next to R136 among the local group, could be 
triggered by the H \emissiontype{I} cloud collision falling down from the stream to the disk, by 
analogy with R136.

The NGC 604 cluster with an age of 3-5 Myr contains more than 200 O-type stars associated 
with a bright H$\alpha$ nebula extending to a radius of 200-400 pc (e.g., \cite{relano09}) at 
a distance of 794 kpc from the Sun \citep{mcconnachie04}. 
The stellar mass of NGC 604 amounts to be $\sim 4 \times 10^5\,M_{\Sol}$ \citep{eldridge11}. 
The feedback of the expanding H \emissiontype{II} region and sequential star formation 
have been discussed from the existence of arc-like H$\alpha$ nebula, warm molecular 
gas and the spatial gradient of the star formation efficiency \citep{young96, tenorio00, tosaki07, 
miura10}. The latter two papers argue that the 2nd generation star clusters are triggered by 
the expansion of the H \emissiontype{II} region excited by the central star cluster. From 
the evolutional stages of molecular clouds, star formation propagates from the central cluster 
where molecular and atomic gas has a cavity to $\sim 100$ pc south where massive Giant 
Molecular Clouds (GMCs) are distributed. The time scale of this propagation is estimated to 
be in the order of $10^6$ yr, comparable to the cluster age \citep{tosaki07}.

In this paper, we focus on the formation mechanism of NGC 604 by investigating the velocity 
structure of H \emissiontype{I} gas around it, particularly in terms of triggering. 
Section 2 describes the archival H \emissiontype{I} data and data analysis. Section 3 gives 
the result of multiple velocity decomposition of the H \emissiontype{I} gas and their distributions. 
We then discuss the possibility of colliding H \emissiontype{I} gas that may trigger molecular 
cloud formation and subsequent massive cluster (NGC 604) formation in Section 4, and 
summarize in Section 5.

\section{Archival data and data analysis}

Entire M33 has been surveyed in atomic gas with Very Large Array (VLA) and in molecular 
gas with the Nobeyama 45m and IRAM 30m telescopes with a spatial resolution better 
than 20$\arcsec$ (corresponding to 77 pc). This enabled us to identify 
individual atomic and molecular clouds, and make statistical studies with their properties, 
such as star-formation activities represented by association of H \emissiontype{II} regions. 

\begin{figure}
\begin{center}
\includegraphics[width=\linewidth]{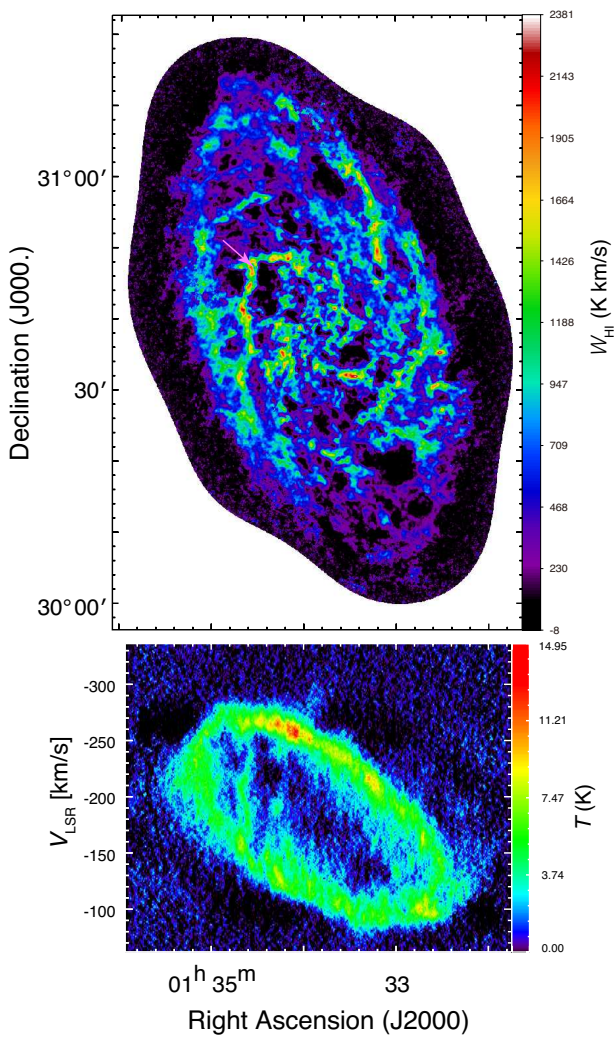} 
 \end{center}
\caption{Top: Integrated intensity map of the VLA H \emissiontype{I} data for the whole M33 disk. 
Bottom: Position-velocity diagram of the H \emissiontype{I} data averaged over the 
declination axis. The arrow denotes the position of NGC 604.}
\label{HI_PV}
\end{figure}

The H \emissiontype{I} mosaic data were obtained from the archive of the VLA with the 
configurations of B, C, and D. They were reduced and imaged by \citet{gratier10} on the 
Common Astronomy Software Applications (CASA). The datasets were 
merged and in the $uv$ plane with subsequent imaging and deconvolution on the merged 
dataset, 4 resolution of 5, 12, 17, and 25 arcsec where obtained by tapering in the UV plane, 
with 1.27 km s$^{-1}$ channel width covering the area of 0.84 deg$^2$ (see \cite{gratier10} 
for more detail). 
Those of 12 arcsec (46 pc) resolution oversampled with 
4 arcsec pixel scale are used in this paper as the signal-to-noise ratio of 
the 5 arcsec map is not sufficient compared to the one at 12 arcsec. 
Fig.~\ref{HI_PV} top is the integrated intensity mosaic map of the H \emissiontype{I} emission 
covering the entire star forming disk of M33, showing dense atomic gas clouds concentrated 
along the spiral arms. Fig.~\ref{HI_PV} bottom shows the position velocity (PV) diagram with 
the intensity averaged along the declination. Because of the inclination and rotation of the galaxy, 
the diagram demonstrates elliptical intensity distribution with the spectra blue-shifted in northeast 
and red-shifted in southwest. 

\begin{figure}
 \begin{center}
 \includegraphics[width=\linewidth]{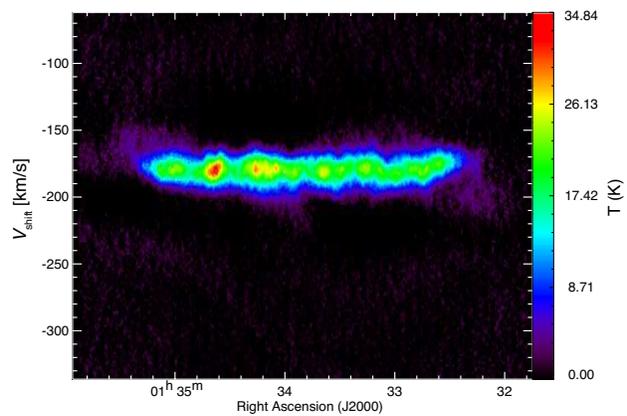} 
 \end{center}
\caption{Position-velocity diagram of the H \emissiontype{I} data averaged over the declination 
axis after shifting the velocity reference by the analytical model for minimizing of the velocity 
gradient due to the rotation and inclination of the galaxy.}
\label{shifted_PV}
\end{figure}

In order to compensate the velocity gradient due to the inclination and galaxy rotation, 
the spectra at each pixel are shifted assuming that the galaxy has a flat rotation following 
the model expressed as a $\tanh$ function as: 
\begin{equation}
V(r) = V_{\infty} \tanh (r/r_0) + V_0
\end{equation}
where $V_{\infty}$ is the circular velocity at $r > r_0$, $r_0$ is the radius where the 
velocity field change from rigid to flat rotation, $V_0$ is the systemic velocity. 
This is basically the same analytical model by \citet{corbelli97} with slightly modified 
parameters to minimized the velocity dispersion of the averaged spectra. 
We first define the rotation center and position angle to have the constant 
velocity centroid along the inclination axis. Then we search for the best rotation model by 
changing these parameters and check the velocity dispersion of the averaged spectra after 
applying the velocity shifts. 
We find that $V_{\infty} = 107.1$ km s$^{-1}$, $r_0 = \timeform{8.27'}$ , and $V_0 = 181.3$ 
km s$^{-1}$ give the best result, while \citet{corbelli97} estimated slightly larger value of 
$r_0 = \timeform{8.47'}$. 
The inclination is compensated by the inclination angle $i = 60\fdg0$, the 
rotation center at (R.A., Dec)$_{J2000}$ =  (\timeform{1h33m51.0s}, \timeform{+30D39'35''}), 
and the position angle $\alpha = 21\fdg0$. Our best fit results give slightly larger $i$ and 
smaller $\alpha$ than previous values given by \citet{regan94} and \citet{paturel03} by a 
few degrees. After compensating the inclination and rotation, the velocity structure is ``flattened" 
as shown in the position velocity diagram of Fig.~\ref{shifted_PV}. The full-width half maximum 
(FWHM) of the averaged spectra of thus shifted data is estimated to be 22.2 km s$^{-1}$. 
In the following analysis, we use the H \emissiontype{I} spectra in thus shifted velocity frame 
$V_{\rm shift}$.

\section{Results}

\subsection{H \emissiontype{I} clouds around NGC 604}

\begin{figure}
\begin{center}
\includegraphics[width=\linewidth]{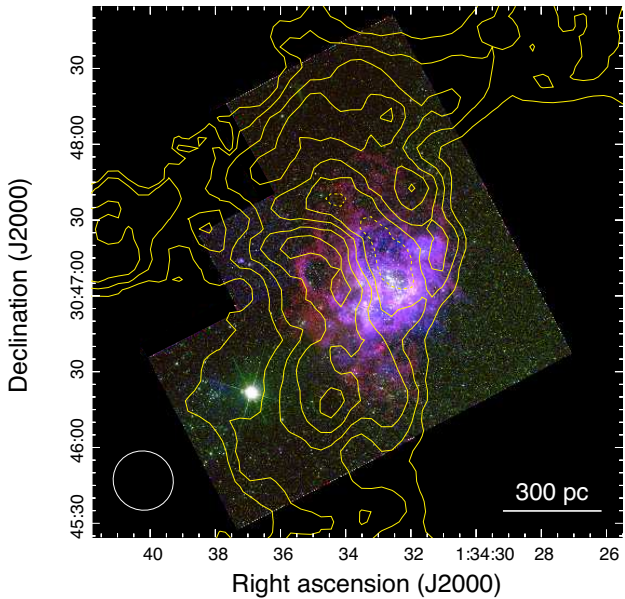} 
 \end{center}
\caption{Composite image of NGC 604 taken with the HST Wide-Field Planetary Camera 2. 
The red, green, and blue colors represent the images of F673N, F547M, and F502N filter 
bands, respectively. 
Overlaid are the contours of the H \emissiontype{I} line integrated intensity starting from 
800 K km s$^{-1}$ with the steps of 300 K km s$^{-1}$. The contours of broken lines 
denote intensity hollows insides. The open circle in the bottom-left corner illustrates the 
synthesized beam size of 12 arcsec.
}
\label{HST}
\end{figure}

\begin{figure*}
 \begin{center}
  \includegraphics[width=\linewidth]{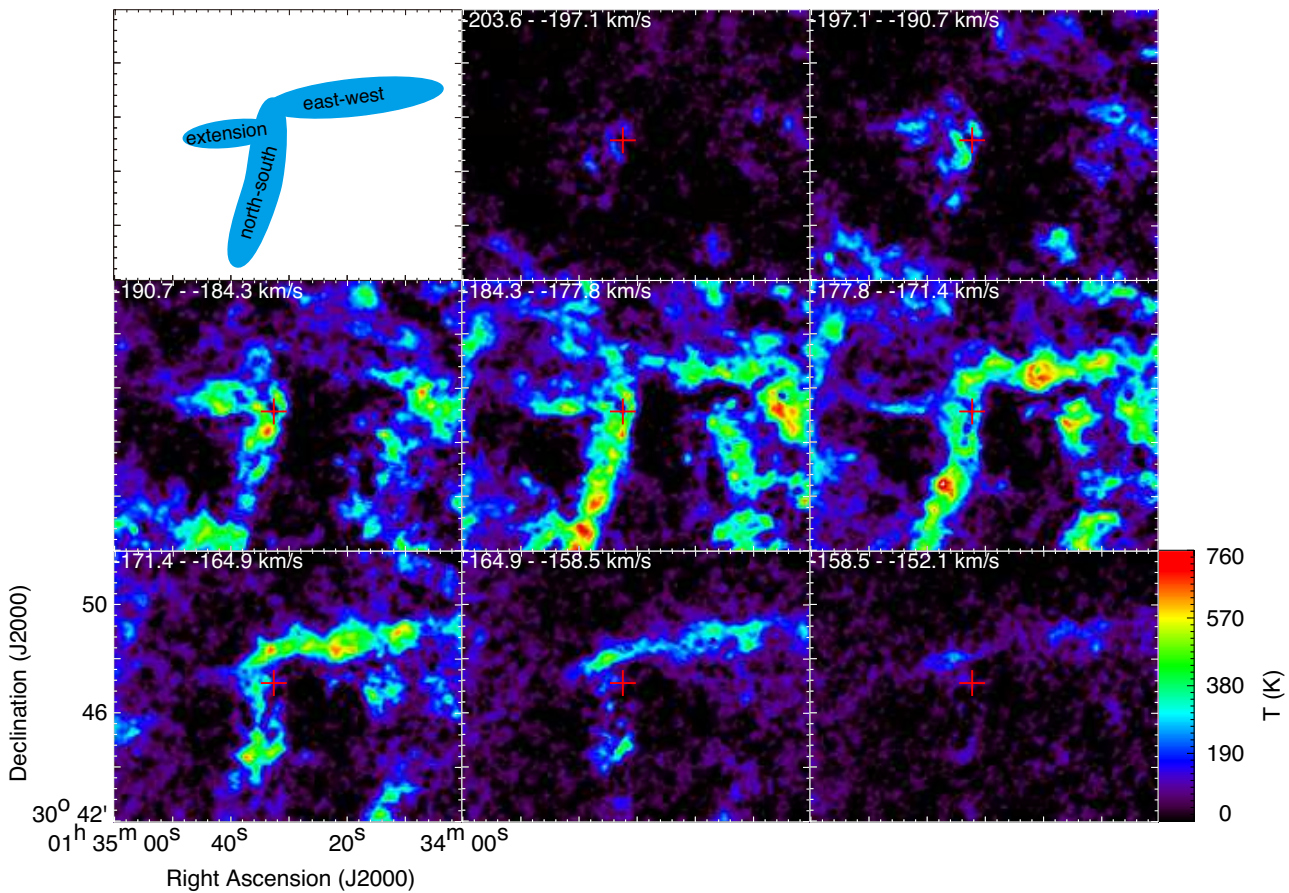} 
 \end{center}
\caption{Velocity channel maps of the clouds around NGC 604 and the associated arm in 
the shifted velocity frame, $V_{\rm shift}$. 
Each map exhibits distributions of the integrated intensity of the H \emissiontype{I} emission 
over the shifted velocity with the integration range designated at the top left of each panel. 
The red cross in each panel denotes the position of NGC 604. 
The top-left panel schematically illustrate the positions of the east-west and north-south 
features and the extension.
}
\label{chmap}
\end{figure*}

\begin{figure}
 \begin{center}
  \includegraphics[width=\linewidth]{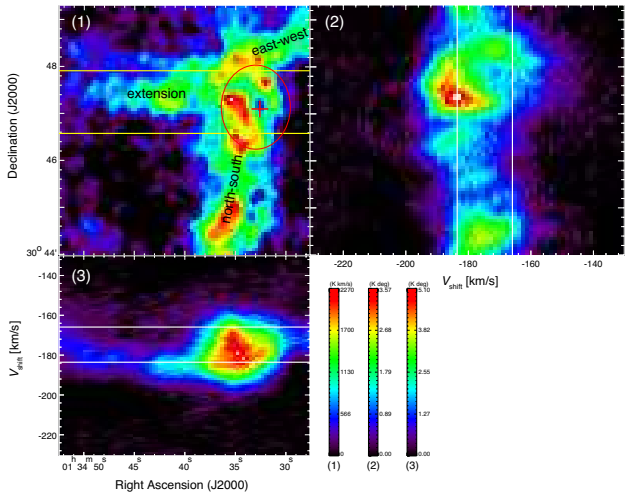} 
 \end{center}
\caption{Zoomed up image of the H \emissiontype{I} integrated intensity map around 
NGC 604 (1), together with the position velocity diagrams averaged over right ascension 
(2) and declination (3). The spectra are shifted in velocity to flattened velocity 
gradient (see text for detail). The red cross denotes the position of NGC 604 with the extent 
of the surrounding H \emissiontype{II} region delineated by the red circle. The 
three color bars for each pane are inserted in the bottom right. 
The two yellow horizontal lines in panel (1) show the integrated range in declination around 
the cavity for Fig.~\ref{cavity}. The white lines in (2) and (3) designate the 
median velocities of the red- and blue-shifted components (see Sec.\ref{decomposition}).
}
\label{N604_PPV}
\end{figure}

Here we focus on spatial and velocity distributions of the H \emissiontype{I} clouds around the 
supergiant H \emissiontype{II} region NGC 604. Fig.~\ref{HST} shows the detailed cloud 
distributions superposed on the optically visible nebulosity of NGC 604 imaged by the 
Hubble Space Telescope (HST). 
The H \emissiontype{I} intensity has a depression seen as a cavity where interstellar gas 
is ionized by the strong UV radiation, and the central cluster is visible through the cavity 
(see the schematic view of \cite{tosaki07}). 
The H \emissiontype{I} clouds associated with the spiral arm in this part of M33 has a bent 
shape with an east-west elongated feature connected with a north-south feature toward NGC 604. 
It is also connected with a feature of short extension stretched to the east by $\sim 700$ pc 
away from the NGC 604 cluster. 
Fig.~\ref{chmap} is the velocity channel map of the H \emissiontype{I} data around NGC 604. 
As seen in these maps and the position-velocity diagram of Fig.~\ref{N604_PPV}, the clouds 
of north-south and east-west features associated with the spiral 
arm are mainly distributed in the velocity range of $-185\ {\rm km\ s^{-1}} 
< V_{\rm shift} < -158\ {\rm km\ s^{-1}}$, whereas the clouds in the extension feature are 
relatively blue-shifted as $-197\ {\rm km\ s^{-1}} < V_{\rm shift} < -170\ {\rm km\ s^{-1}}$. 
They both have, however, complex spatial and velocity distributions and it is difficult to 
separate them simply by a threshold velocity.

\begin{figure}
 \begin{center}
  \includegraphics[width=\linewidth]{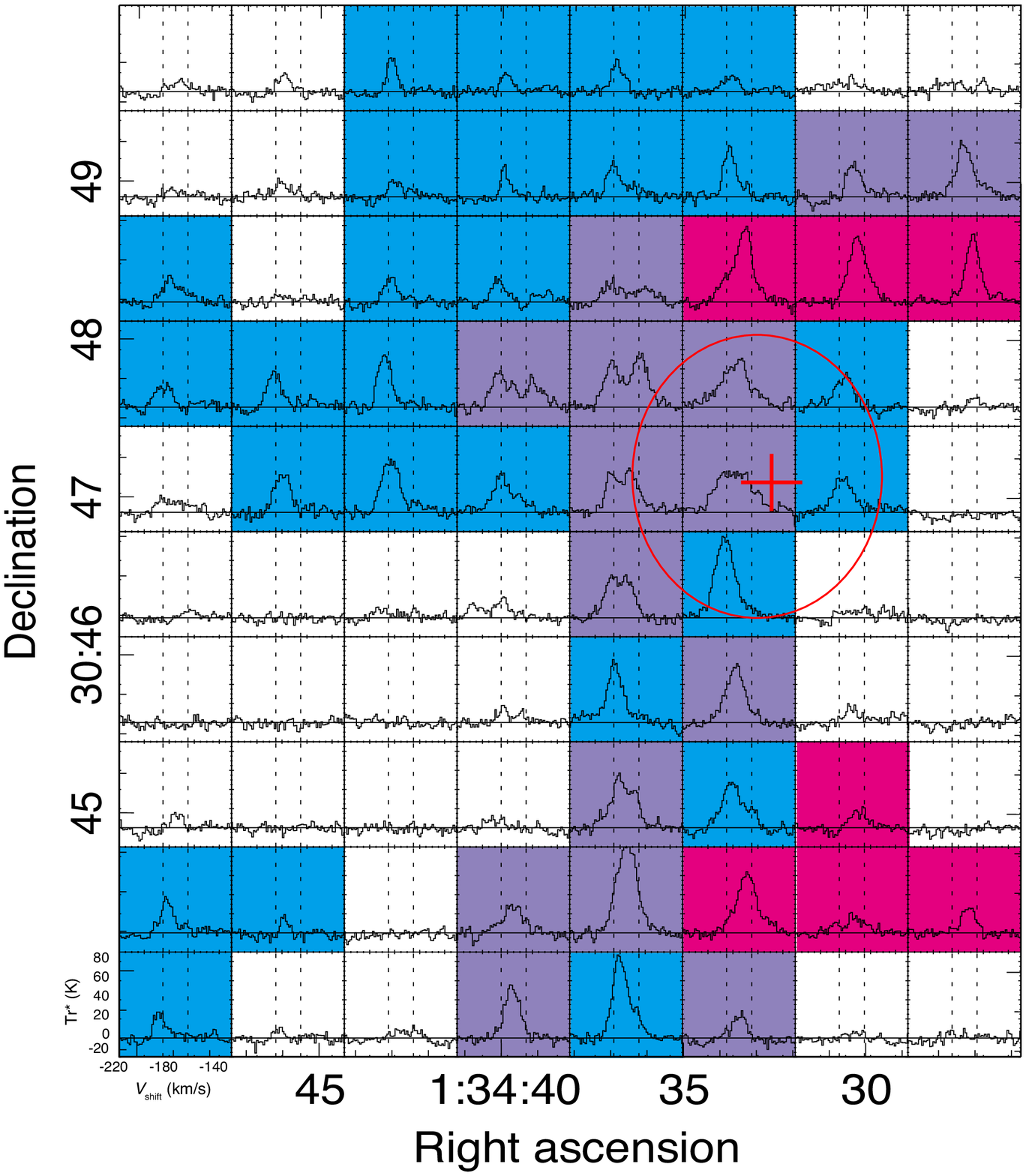} 
 \end{center}
\caption{The H \emissiontype{I} profile map around the NGC 604 region. The spectra are 
shifted in velocity according to the analytic model (see text), and spatially rebinned down to 
20 arcsec, and shown the grid separation of 40 arcsec. 
The red cross denotes the position of NGC 604 with the extent 
of the surrounding H \emissiontype{II} region delineated by the red circle. 
The 2 dotted lines designate $V_{\rm shift} = -185$ and $-165$ km s$^{-1}$, the initial values 
of the central velocities for the Gaussian fittings. 
The background color of cyan and magenta denote that the spectra are dominated by 
blue and red-shifted single component, respectively, while that of purple denote that 
spectra are clearly double-peaked, or single-peaked with intermediate peak velocity.
}
\label{pfmap}
\end{figure}

Fig.~\ref{pfmap} is an H \emissiontype{I} profile map with the spatial resolution of $20\arcsec$ 
(after 25 pixel binning) and $40\arcsec$ separation each around NGC 604. 
As clearly seen in the velocity distribution and the double-peaked spectral shapes, the 
H \emissiontype{I} gas consists of at least two discrete velocity components. 
The blue-shifted component is dominated by the emission at the velocity of $V_{\rm shift} 
\sim -185$ km s$^{-1}$, while the red-shifted one is at $V_{\rm shift} \sim -165$ km s$^{-1}$ 
although the entire cloud has complex velocity distribution with spectra on intermediate velocity 
and merged spectra at some positions. 
The background color in cyan and magenta denote that the H\emissiontype{I} spectra is 
dominated by blue- and red-shifted components, respectively, with single-peak velocity close 
to $-185$ km s$^{-1}$ and $-165$ km s$^{-1}$, respectively. 
The purple one, on the other hand, show that the spectra imply the existence of mixed 
H\emissiontype{I} gas as represented by double-peaked profile or single-peaked one with 
the peak velocity of the intermediate values. 

Note that the clouds in the extension part are dominated with the blue-shifted gas, whereas 
the red-shifted component is strong in the western parts of the clouds. 
The spectra within the NGC 604 H \emissiontype{II} region delineated by the blue ellipse 
in Fig.~\ref{pfmap} have relatively large velocity dispersion, where the double-peaked profiles 
are clearly detected outside the ellipse, implying that the complex velocity structure is not 
only due to the feedback effect of the central cluster, but rather large-scale gas merging of 
the two velocity components. 

\subsection{Two-component Gaussian decomposition}
\label{decomposition}

\begin{figure}
 \begin{center}
  \includegraphics[width=\linewidth]{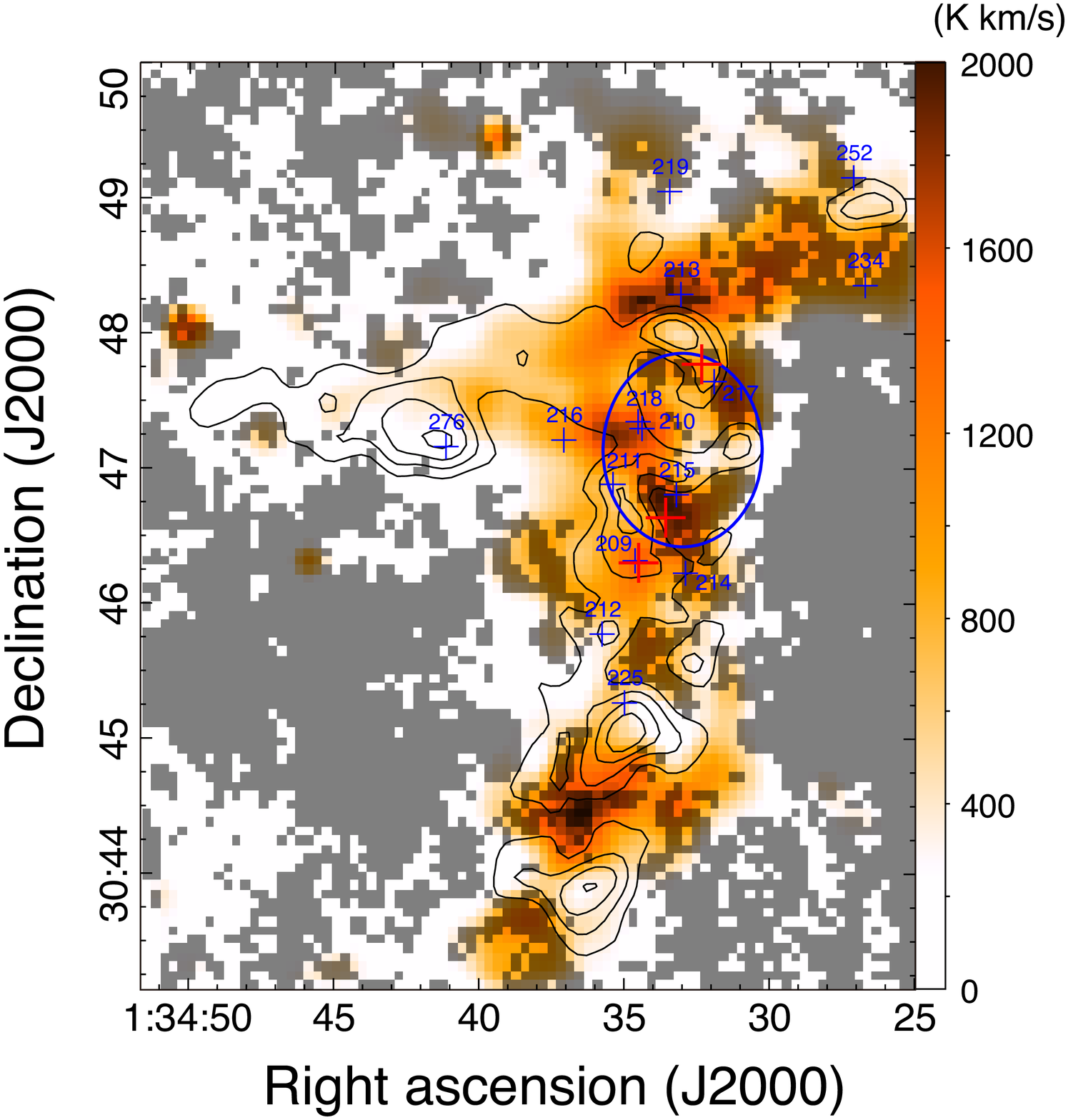} 
 \end{center}
\caption{The image represents the total intensity of the red-shifted ($V_{\rm shift} \approx 
-165$ km s$^{-1}$) H \emissiontype{I} component derived from the Gaussian fitting, while 
the overlaid contours are that of the blue-shifted ($V_{\rm shift} \approx -185$ km s$^{-1}$) 
component. The intensity is calculated from the area of the fitted Gaussian function. 
The image and contours are smoothed by 3-pixel Gaussian filters. The contour levels are 
from 500 K km s$^{-1}$ with the steps of 250 K km s$^{-1}$. 
The gray shaded areas denote the pixels where 2-component Gaussian fitting fails.
The blue ellipse delineates the extent of the optical nebula of NGC 604. 
The blue crosses designate the positions of GMCs with their numbers identified in CO 
$J=$2--1 by \citet{gratier12}. The red crosses show the peak positions of CO $J=$1--0 
intensity (see text and Fig.~\ref{COdist}.)
}
\label{degauss}
\end{figure}

In order to characterize physical propertied of the H \emissiontype{I} clouds around NGC 604, 
decomposition of the spectra by two-component Gaussian fittings are attempted. For the 
spectrum at each pixel, least-squares regression fittings are adapted with the two-component 
Gaussian functions. The initial values of the central velocity for the fittings are given as 
$V_{\rm shift} = -185$ km s$^{-1}$ and $-165$ km s$^{-1}$, and the peak intensities and 
velocity dispersions are set to be free parameters. After the iterations, the result of the fitting 
for each pixel is evaluated, and in case the fitting failed, fitting is repeated with the single 
component Gaussian function. 
Fig.~\ref{degauss} exhibits the distributions of the separated two velocity components 
with the $-165$ km s$^{-1}$ (red-shifted) component as the image and the $-185$ km s$^{-1}$ 
(blue-shifted) one as the contours. Note that when the spectrum is blended, the fitted 
single-component Gaussian is attributed to the red-shifted component, and the gray 
shaded area denotes the pixels where two-component Gaussian fittings failed. 
The median values of the resultant central velocity for the red- and blue-shifted components 
are $-165.91$ km s$^{-1}$ and $-183.57$ km s$^{-1}$, respectively. 
The red-shifted component distributes mainly along the arm, while the blue-shifted component 
spread from the extension to the region near NGC 604 and also further down to the south 
along the arm. Toward the central cluster of NGC 604 H \emissiontype{II} region, both of the 
two velocity components exhibits depression. Along the arm between \timeform{+30D43'30"} 
$<$ Dec.~$<$ \timeform{+30D45'30''}, the H \emissiontype{I} cloud (hereafter the southern cloud) 
consist of the two velocity components, while they are mixed at the middle part (Dec.~$\approx 
\timeform{+30D44'30"}$) of the cloud.

The H \emissiontype{I} column density and total mass of thus separated components are 
derived. The column density of H \emissiontype{I}, $N_{\rm H \emissiontype{I}}$, is calculated 
from the radiative transfer equation with the integrated intensity of 
H \emissiontype{I}, $W_{\rm H \emissiontype{I}}$ for each pixel assuming the optically thin 
condition as:
\begin{equation}
N_{\rm H \emissiontype{I}} = 1.823 \times 10^{18} \times W_{\rm H \emissiontype{I}} 
\hspace{1cm} {\rm [cm^{-2}]}\ , 
\end{equation}
where $W_{\rm H \emissiontype{I}}$ is the integrated intensity of the 
H \emissiontype{I} emission line given as $W_{\rm H \emissiontype{I}} = \int T_{\rm b} dv$, 
with the observed H \emissiontype{I} brightness temperature $T_{\rm b}$.
The blue-shifted cloud has a peak column density of $N_{\rm H \emissiontype{I}} = 
2.3 \times 10^{21}$ cm$^{-2}$ 
at (R.A., Dec.)$_{J2000}$ = (\timeform{1h34m41.3s}, \timeform{+30D47'14"}) for the extension, 
and $N_{\rm H \emissiontype{I}} = 2.7 \times 10^{21}$ cm$^{-2}$ at (\timeform{1h34m35.0s}, 
\timeform{+30D45'2''}) for the southern cloud. The atomic total mass of the blue-shifted clouds 
is estimated by summing up the column density of each pixel to be $6 \times 10^6~M_{\Sol}$. 
The red-shifted cloud has, on the other hand, peak column density of 
$N_{\rm H \emissiontype{I}} = 3.4 \times 10^{21}$ cm$^{-2}$ near the border of the NGC 604 
H \emissiontype{II} region. The atomic total mass of the red-shifted clouds is $9 \times 
10^6~M_{\Sol}$. 

\subsection{Associated CO clouds and their star formation activities}

\citet{tosaki11} surveyed entire disk of M33 for molecular clouds in CO $J=$1--0 emission 
with the Nobeyama 45m telescope, and revealed molecular fraction $f_{\rm mol}$ distribution 
as a function of the radial distance from the galactic center. The azimuthally averaged 
distribution demonstrates a trend that 
higher $f_{\rm mol}$ for the inner 1.5 kpc region with a peak value of $\sim 0.2$, while 
the outer part has typical $f_{\rm mol}$ of $\lesssim 0.04$. They suggest higher metallicity 
for the inner part of the galaxy. This molecular fraction is given by a ratio of molecular and 
atomic surface densities as $f_{\rm mol} = \Sigma_{\rm H_2} / 
(\Sigma_{\rm H \emissiontype{I}} + \Sigma_{\rm H_2})$, and hence directly compared 
with a molecular to atomic mass fraction of $M_{\rm H_2} / (M_{\rm H \emissiontype{I}} 
+ M_{\rm H_2})$ if the molecular and atomic clouds are sampled from the same area. 
For the clouds in the present region around NGC 604 displayed in 
Fig.\ref{degauss}, the total atomic gas mass is estimated by summing up the values of 
$ N_{\rm H \emissiontype{I}}$ and multiplying with the area of the grid and the proton mass to be 
$M_{\rm H \emissiontype{I}} \sim 1.5 \times 10^7\ M_{\Sol}$, while the molecular cloud 
mass in the same area is estimated similarly from molecular column density 
$N_{\rm H_2}$ to be $M_{\rm H_2} = 2 \times 10^6\ M_{\Sol}$ from the CO $J=$1--0 data 
with an assumption of an $X_{\rm CO}$ factor of $3 \times 10^{20}\ {\rm cm^{-2} 
(K~km~s^{-1})^{-1}}$ \citep{wilson90}. 
This gives a $f_{\rm mol}$ ratio of $\sim 0.12$ for the NGC 604 region, significantly higher 
than the averaged values ($\sim 0.04$) estimated for the galactic radius of $\sim 2.8$ kpc 
in M33 by \citet{tosaki11}.

On the other and, \citet{gratier12} entirely surveyed the galaxy in the CO ($J=$2--1) line 
with the IRAM 30m telescope and identified 337 GMCs using the CPROPS (Cloud PROPertieS) algorithm \citep{rosolowsky06}, and compared them with the 
same dataset of H \emissiontype{I} as the present work.  
In the region of our interest around NGC 604, there are GMC \#209-\#219 and \#225. 
GMC \#276 corresponds to the peak of the blue-shifted clouds in the extension feature, 
while GMC \#225 corresponds to the blue-shifted peak of the southern cloud. 
In and around the area of the NGC 604 H \emissiontype{II} region, several GMCs are 
distributed (Fig.~\ref{degauss}). 

\citet{gratier12} classified the GMCs into 3 types by the star formation activities, i.e., type A clouds 
are non-star-forming clouds without any indication of associated stars, type B clouds have 
embedded star formation with 8 and 24 $\mu$m emission but not seen in H$\alpha$ and FUV, 
and type C clouds have exposed star formation with detections in all these bands, 
similar manner to \citet{fukui99} and \citet{kawamura09} for those in the LMC. 
This classification is made by several testers independently, and the judgement is not 
necessarily unique but diverse for some GMCs. However, majority classification gives an 
idea of star formation activities of the clouds. 
The CO emission of GMC \#276 has a peak velocity corresponding to the blue-shifted 
H \emissiontype{I} component and is classified as type C. GMC \#225 and \#216 also belong 
to the blue-shifted clouds and the major classifications are type C, while another blue-shifted 
GMC \#212 is classified as type A. On the other hand, GMC \#209 is a red-shifted type A cloud, 
and GMC \#211 a red-shifted type C cloud. GMC \#210 and \#218 have double peaked CO 
profiles and classified as type A or B. GMC \#213 (type B/C), \#214 (type A), \#215 (type C), 
and \#217 (type B) have H \emissiontype{I} and CO lines of blended velocity components. 
Both the blue- or red-shifted clouds have a variety of properties in star formation.
There is no clear difference in the star formation trend between the blue and 
red-shifted components. In the present region, about one quarter (27\%) of the clouds are 
attributed to Type A. This is slightly large fraction compared to the statistics of the whole 
clouds sampled from the entire galaxy (15\%).

\section{Discussion}

\subsection{Kinematics of the H \emissiontype{I} clouds}

\begin{figure}
 \begin{center}
  \includegraphics[width=\linewidth]{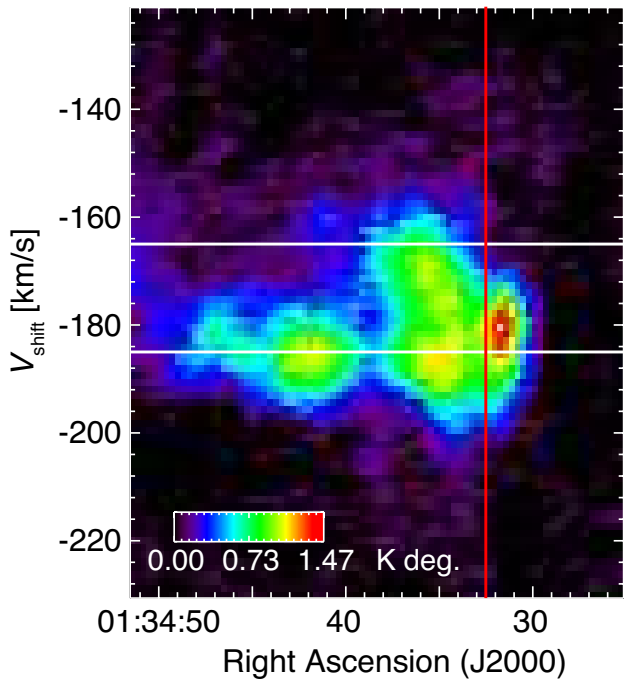} 
 \end{center}
\caption{Position-Velocity diagram toward the region including the H \emissiontype{I} 
cavity around NGC 604 and the extension cloud. The H \emissiontype{I} spectra are integrated 
over the axis of declination for the range between the 2 white lines in Fig.~\ref{N604_PPV}. 
The red vertical bar designates the position of H \emissiontype{I} cavity centered at the 
NGC 604 super star cluster, while the horizontal white ones are the median velocities of 
the decomposed red- and blue-shifted components.
}
\label{cavity}
\end{figure}

As shown above, the H \emissiontype{I} clouds around NGC 604 have complex velocity 
distributions, and they consist of at least two discrete velocity components. Toward some 
positions, the spectra appear to be merged, having peak velocity between the two components.  
We consider the origin of the velocity structures. 
One possibility is the expanding motion energized by the feedback effect of the super 
star cluster NGC 604, which is surrounded by multiple shell-like structures of the H$\alpha$ 
nebulosity. \citet{tosaki07} and \citet{miura10} discussed these phenomena from molecular 
data by single-dish and interferometric observations, respectively. The CO clouds 
exhibits indications of the interaction with the cluster as high $J=$3--2/1--0 line ratio 
($R_{3-2/1-0}$) along the H$\alpha$ shell, implying the interaction of the molecular 
clouds with the expanding H \emissiontype{II} region. The Star Formation Efficiency 
(SFE) shows decreasing trend with the distance from the NGC 604 cluster, implying the 
sequential star formation triggered by the feedback effects. The molecular clouds around 
NGC 604 have not been reported to exhibit expanding motion due to the feedback. 
Fig.~\ref{cavity} is the PV diagram focused on the cavity of the H \emissiontype{I} cloud and 
the extension cloud to the east (see Fig.~\ref{N604_PPV}). Toward the H \emissiontype{I} 
cavity, one can clearly recognize the depression of H \emissiontype{I} intensity. This 
is a peace of evidence of feedback by NGC 604 affecting on the H \emissiontype{I} cloud, 
likely as a photo-ionization. However, the velocity field does not appear to have a simple 
expanding motion in this PV diagram. 
To the west of the cavity (R.A.\ $\sim \timeform{1h34m32s}$), the H \emissiontype{I} spectra 
appear to be single peaked, while those to the east (R.A.\ $\sim \timeform{1h34m35s}$) 
are multi-peaked (with blue- and red-shifted components) or having large velocity dispersions. 
The eastern extension cloud has blue-shifted velocity and connected to the blue-shifted 
component around the H \emissiontype{I} cavity. In addition, the distributions of the 2 velocity 
components have much larger extent than the H$\alpha$ nebulosity. 
If we make a simple assumption that the blue- and red-shifted H \emissiontype{I}  clouds 
are due to the expanding motion with the velocity of 10 km s$^{-1}$, the kinetic energy 
of the clouds are estimated to be as much as $6 \times 10^{51}$ erg. 
We thus conclude that the feedback effect of NGC 604 on the complex velocity structures 
of the H \emissiontype{I} clouds is only limited for small surrounding area. 
If we consider kinetic energy injection by supernova (SN) explosions with $\sim 5\%$ 
efficiency \citep{kruijssen12}, it requires $\sim 100$ SNe within the age of NGC 604 cluster. 
The number is still much larger than that expected from the cluster scale. 
Remarkable energy of radiation is released from the massive stars. 
The radiation pressure is suggested to evacuate the interstellar medium from the galactic 
disk as the galactic winds if the cluster mass is more than $\sim 10^6\,M_{\Sol}$, while the 
bubble radius is limited to be less than $\sim 30$ pc for the cluster mass of $3 \times 10^5 
M_{\Sol}$ \citep{murray11}.

As an alternative possibility to explain the velocity fields, we propose a gas accretion 
scenario onto the spiral arm of the disk, that triggered active star formation of NGC 604. 
Similar to the case of the Magellanic bridge and the Magellanic stream, M31-M33 system 
is known to have diffuse gas bridge connecting the 2 galaxies \citep{braun04, putman09, 
lockman12}. \citet{grossi08} detected H \emissiontype{I} clouds in the proxy of M33, 
and suggest that they are HVCs accreting and fueling the M33 disk for star formation. 
If this is the case, the formation of NGC 604 is supposed to be triggered by the collision 
of infalling H \emissiontype{I} clouds, similar to the case for R136 in the LMC. 
The velocity separation between the 2 components is $\sim 50$ km s$^{-1}$ for R136, 
about a factor of 2-3 larger than the present case of NGC 604. 
\citet{bekki08} demonstrated by numerical simulations that M31 and M33 had a close 
encounter at 4-8 Gyr ago, that resulted in forming gas streamer stretched by tidal force. 
\citet{vdmarel12} predicted future encounters and possible mergers of 
M31, M33, and the Milky way in several Gyr. 
The orbital motion is, however, very complicated and also depend on the assumed initial 
conditions, particularly at the region close to the galactic disk. It is, therefore 
difficult to expect the relative velocity of the infalling gas with respect to the galactic disk.  
The blue-shifted H \emissiontype{I} clouds dominating the extension feature is, however, 
candidate infalling gas from the morphological characteristics. 
The impact of the collisions between the H \emissiontype{I} clouds seem to be 
significant, but the bulk energy of the collision is hard to be estimated. 

\subsection{Formation of molecular gas by shock triggering}

\begin{figure}
 \begin{center}
  \includegraphics[width=\linewidth]{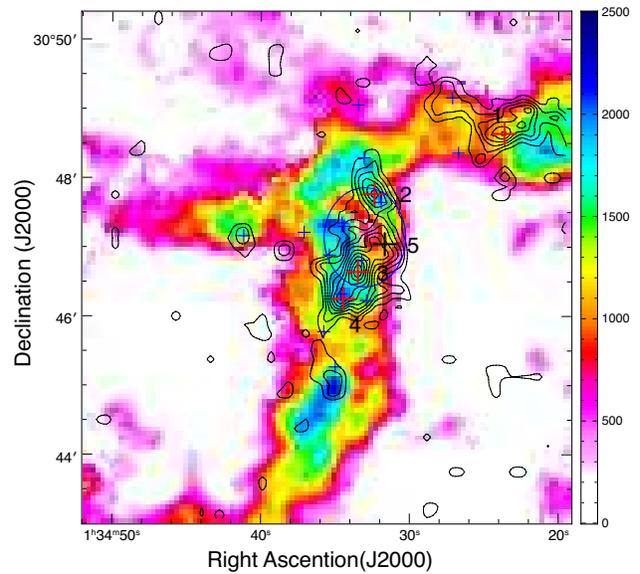} 
 \end{center}
\caption{Integrated intensity map of H \emissiontype{I} shown by the pseudo color overlaid 
with the contours of CO $J=$1--0 integrated intensity \citep{tosaki07}. The crosses denote 
the positions whose CO and H \emissiontype{I} spectra are displayed in Fig.~\ref{CO-HIsp}. 
The four red crosses correspond to the positions of CO intensity peaks, while the black one to 
the NGC 604 cluster. The blue crosses are the positions of GMCs defined in 
CO $J=$2--1 by \citet{gratier12}. 
The contours are from 0.6 K km s$^{-1}$ with steps of 0.9 K km s$^{-1}$.
The blue ellipse delineates the extent of the optical nebula of NGC 604.
}
\label{COdist}
\end{figure}

\begin{figure}
 \begin{center}
  \includegraphics[width=\linewidth]{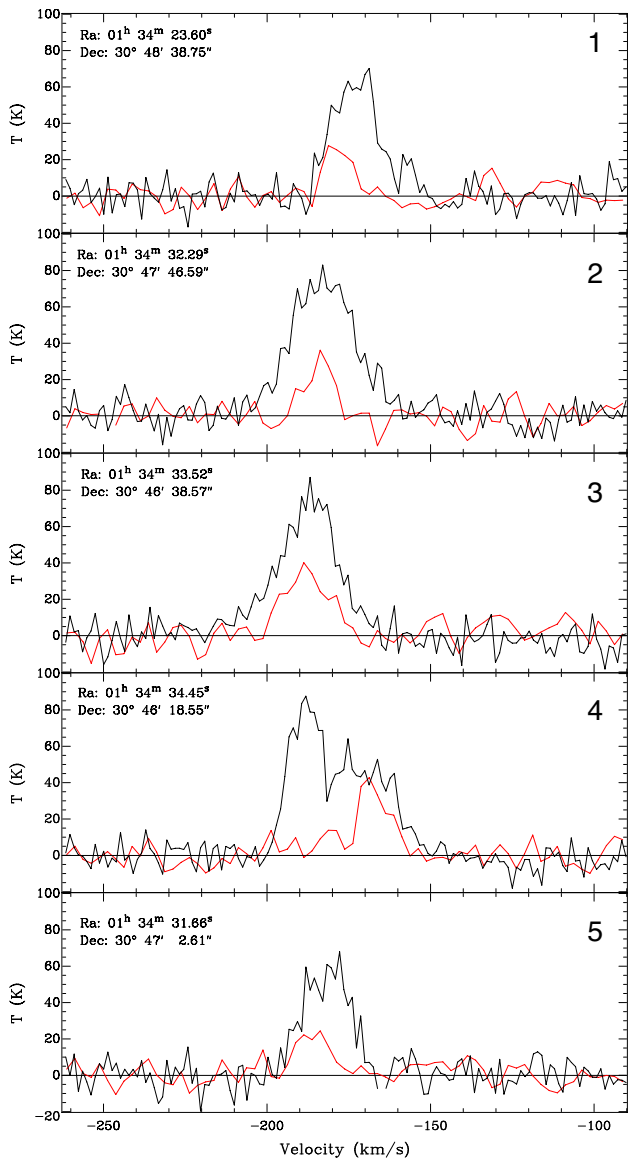} 
 \end{center}
\caption{H \emissiontype{I} (black) and CO (red) spectra at the 5 positions shown in 
Fig.\ref{COdist}. The CO spectra are scaled by a factor of 5.
}
\label{CO-HIsp}
\end{figure}

Massive molecular clouds are distributed around NGC 604 where active star formation is ongoing. 
Fig.~\ref{COdist} reveals the integrated intensity distribution of CO $J=$1--0 by contours 
\citep{tosaki11} overlaid on the H \emissiontype{I} integrated intensity map. The CO clouds 
show concentrations in a region around the H \emissiontype{I} cavity of NGC 604 with 
clumpy shapes peaked at 3 positions denoted by the red crosses. The distributions of the 
$J=$2--1 line intensity, on the other hand, are slightly different, namely that GMC \#215 and 
\#217 have peak positions closer to the central cluster of NGC 604 than the $J=$1--0 peak 2 
and peak 3, while the position of peak 4 corresponds well to that of GMC \#209 (see 
Fig.~\ref{degauss}). There is another CO cloud with peak 1 on the east-west feature to the 
north of NGC 604 (hereafter the northern cloud), where H \emissiontype{I} intensity is 
depressed as a complementary distribution with CO. 
Fig.\ref{CO-HIsp} illustrates the H \emissiontype{I} and CO spectra sampled at the 3 peak 
positions of CO $J=$1--0 together with those at the H \emissiontype{I} cavity edge. 
At the peak 2 and 3, CO emission lines are detected at the intermediate velocity of the two 
H \emissiontype{I} components, while at peak 4 the CO line is strong for the red-shifted 
component. The CO clouds are, on the other hand, detected also from the blue-shifted 
component at some positions around the H \emissiontype{I} cavity such as position 5. 
Another CO cloud with peak 1, the CO emission has relatively blue-shifted velocity with 
respect to the H \emissiontype{I} line. These imply that not only H \emissiontype{I}, but 
the CO emission also has complex velocity fields. 

\begin{figure}
 \begin{center}
  \includegraphics[width=\linewidth]{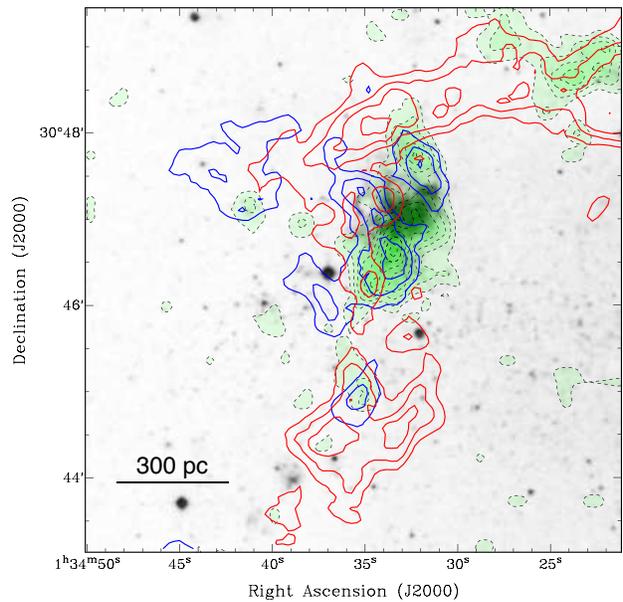} 
 \end{center}
\caption{Distributions of the CO clouds is demonstrated by the green shaded area on top of 
the optical image of DSS-red in gray scale. Overlaid are the blue and red contours representing 
the blue- and red-shifted H \emissiontype{I} clouds with the integrated ranges of $-214\ {\rm 
km\ s^{-1}} < V_{\rm shift} < -184\ {\rm km\ s^{-1}}$ and $-175\ {\rm km\ s^{-1}} < V_{\rm shift} 
< -143\ {\rm km\ s^{-1}}$, respectively.
}
\label{COrenzo}
\end{figure}

Fig.~\ref{COrenzo} demonstrates the CO cloud distributions compared with those of the 
red- and blue-shifted H \emissiontype{I} clouds represented by the contours whose 
integrated velocity ranges are $-214\ {\rm km\ s^{-1}} < V_{\rm shift} < -184\ {\rm km\ s^{-1}}$ 
and $-175\ {\rm km\ s^{-1}} < V_{\rm shift} < -143\ {\rm km\ s^{-1}}$, respectively. 
The CO emission is detected toward the region where the blue- and red-components are 
overlapping around NGC 604 and in the southern cloud. The northern cloud, on the other hand, 
does not have strong blue-shifted H \emissiontype{I} emission in this velocity range, but the 
Gaussian decomposition succeeded to separate the blue-shifted component near GMC \#252 
(see Fig.~\ref{degauss}). 

As discussed above, it is suggested that the 2 velocity components of the H \emissiontype{I} 
clouds seem to be due to the gas infall originated from the tidal interaction between M31 and 
M33, and the collision of the H \emissiontype{I} clouds triggered the formation of NGC 604 
super star cluster. If this is the case, the idea of molecular cloud formation by the cloud-cloud 
collision should be considerable. Numerical simulations of colliding gas flow and subsequent 
thermal instabilities have been investigated by many authors \citep{field69, wolfire95, 
koyama00, koyama02, hennebelle07, inoue12}.

The dynamical crossing time of the two H \emissiontype{I} clouds can be calculated from 
the size of the clouds divided by the relative collision velocity. If we assume the size 
of the colliding system as $\sim 600$ pc, the collision velocity of $\sim 20$ km s$^{-1}$, 
it gives the crossing time of $\sim 3 \times 10^7$ yrs. Within this period, molecular gas 
formation is feasible if the density is high enough \citep{goldsmith07}. Therefore it is 
suggested that collisions of H \emissiontype{I} clouds could enhance the local molecular 
fraction of the interstellar gas, and accelerate the chemical evolution of the galaxy. 
The local enhancement of $f_{\rm mol} \sim 0.12$ derived above could be affected by 
the gas ionization by the UV radiation from the massive star members of NGC 604, 
namely that the surrounding atomic gas is ionized by the cluster, while the inner molecular 
clouds are more shielded from the UV irradiation due to the higher column density. 
The molecular clouds have, however, much more extended beyond the extent of the 
optically visible H \emissiontype{II} nebula. The effect of UV ionization seems to be limited, 
and we hence suggest molecular formation induced by the shock cloud compression 
triggered by the H \emissiontype{I} cloud collisions.

\subsection{Large-scale gas fueling by M31-M33 tidal interaction and triggering of 
supermassive clusters}

As discussed above, the formations of molecular clouds and subsequently NGC 604 are 
likely to be triggered by collisions of H \emissiontype{I} clouds. We suggest that 
these triggering is not an unique phenomenon for NGC 604, but works also for other 
giant H \emissiontype{II} regions such as NGC 595, where we see similar double-peaked 
H \emissiontype{I} profiles. As suggested by \citet{grossi08}, continuous mass accretion 
fuel the galaxy with fresh gas, that keeps active star formation of M33. 
They identified H \emissiontype{I} clouds with high relative velocity in the periphery of M33, 
and a blue-shifted cloud named AA16 in their catalog is located $\sim 3$ kpc away from 
NGC 604 to the east. This might be an indication of gas infall from galaxy halo or further 
outside. The origin of the infalling gas is unknown, but the H \emissiontype{I} stream between 
M31 and M33 gives a hit of tidal interaction, similar to the formation scenario of the 
Magellanic stream. Such close encounters of galaxies, or mergers and collisions, are 
suggested to be common triggering phenomena for massive star cluster formation in the 
evolutional sequence of galaxies.

\section{Summary}

The H \emissiontype{I} archival data obtained with VLA has been reanalysis, and 
their velocity structures have been investigated. Particularly for the region around 
the super star cluster NGC 604. The results are summarized below.
\begin{itemize}
\item The velocity field of the H \emissiontype{I} clouds around the super star cluster 
NGC 604 appear to have complex structures containing multiple velocity components 
indicated by double-peaked spectral profiles. 
\item They are decomposed by fittings to the two component Gaussian functions, 
whose central velocities are typically separated by 20 km s$^{-1}$.
\item Thus separated two velocity components are distributed far beyond the extent 
of the optically visible H \emissiontype{II} region of NGC 604, and they do not show 
clear expanding velocity feature.
\item The atomic gas mass of each blue- and red-shifted component are estimated 
to be $6 \times 10^6\ M_{\Sol}$ and $9 \times 10^6\ M_{\Sol}$, respectively. 
\item Molecular clouds identified by CO observations are associated with H \emissiontype{I} 
clouds of both velocity components, and the CO clouds are mainly distributed 
towards the overlapped regions of the 2 velocity H \emissiontype{I} components.
\item It is suggested that the H \emissiontype{I} clouds are infalling onto the 
galactic disk and collision of the clouds induced the molecular cloud formation in 
$\sim$ a few $\times 10^7$ yrs.
\item As the mass accretion continues, the collisions of the H \emissiontype{I} clouds 
trigger the formation of super-massive star cluster of NGC 604 in several Myr ago.
\item As suggested for the formation mechanism of R136 in the LMC, 
colliding H \emissiontype{I} gas originated from the past close encounter and the 
tidal effect between galaxies can be a major trigger of super star cluster formation. 
This is consistent with the galaxy merger paradigm for the galaxy evolution scenario.
\end{itemize}

\section{Funding}
The work is supported by Japan Society for the Promotion of Science (JSPS) 
KAKENHI Grant Number JP15H05694. 

\begin{ack}
Based on observations made with the NASA/ESA Hubble Space Telescope, and obtained from the Hubble Legacy Archive, which is a collaboration between the Space Telescope Science Institute (STScI/NASA), the Space Telescope European Coordinating Facility (ST-ECF/ESA) and the Canadian Astronomy Data Centre (CADC/NRC/CSA).
The National Radio Astronomy Observatory is a facility of the National Science Foundation operated under cooperative agreement by Associated Universities, Inc.
The Nobeyama 45-m radio telescope is operated by Nobeyama Radio Observatory, a branch of National Astronomical Observatory of Japan.

\end{ack}

\end{document}